\documentclass[fleqn,usenatbib]{mnras}

\usepackage{newtxtext,newtxmath}
\usepackage[ampersand]{easylist}
\usepackage{ulem}

\usepackage[T1]{fontenc}
\usepackage{ae,aecompl}

\usepackage{graphicx}	
\usepackage{amsmath}	
\usepackage{amssymb}	
\usepackage{comment}

\newcommand{\Msun}{\ensuremath{{\rm M}_{\odot}}}

\newcommand{\Ks}{$K_s$}

\newcommand{\gzK}{$gzK_s$}
\newcommand{\PEgzK}{$P\!E$-$gzK_s$}
\newcommand{\SFgzK}{$SF\!$-$gzK_s$}
\newcommand{\zs}{$z$$\sim$}

\newcommand{\Mstars}{$M_{\star}$}

\newcommand{\Mumpeg}{$M_{\star}^{{\text{UMPEG}}}$}
\newcommand{\Msat}{$M_{\star}^{\text{sat}}$}

\title[LARgE Survey -- III. Environments of $z \sim 1.6$ UMPEGs]{LARgE Survey -- III. Environments  of Ultra-Massive Passive Galaxies at Cosmic Noon:  BCG progenitors growing through mergers}

\author[M.~Sawicki et al.]
{\hspace{-1.7mm}
Marcin Sawicki$^{1, 2}$\thanks{E-mail: {\tt marcin.sawicki@smu.ca}}\thanks{Canada Research Chair},
Liz Arcila-Osejo$^{1}$, 
Anneya Golob$^{1}$,
Thibaud Moutard$^{1}$,
\newauthor
St\'ephane Arnouts$^{3}$,
Gurpreet Kaur Cheema$^{1}$
\\$^{1}$Department of Astronomy \& Physics and the Institute for Computational Astrophysics, Saint Mary's University, 923 Robie Street, Halifax,\\ Nova Scotia, B3H 3C3, Canada
\\$^{2}$Herzberg Astronomy and Astrophysics, National Research Council of Canada, 5071 West Saanich Rd., Victoria, BC V9E 2E7, Canada
\\$^{3}$Laboratoire d'Astrophysique de Marseille, 38 rue Frederic Joliot Curie, Universit\'e Aix-Marseille, Marseille, F-13388, France
}

\date{MNRAS, in press}

\pubyear{2020}

\begin{document}
\label{firstpage}
\pagerange{\pageref{firstpage}--\pageref{lastpage}}
\maketitle
\begin{abstract}
We study the environments of a sample of 61 extremely rare \zs1.6\ Ultra-Massive Passively Evolving Galaxies (UMPEGs: stellar masses \Mstars~$>10^{11.5}$\Msun) which -- based on clustering analysis 
presented in \citet{Cheema2020}
 -- appear to be associated with 
very massive (M$_{\rm halo}\sim$10$^{14.1}h^{-1}$~M$_\odot$)
dark matter halos that are 
likely to be
the progenitors of \zs0 massive (Coma- and Virgo-like) galaxy clusters.  We find that UMPEGs on average have fewer than one satellite galaxy with mass ratio \Msat~:\Mumpeg $\geq$1:5 (i.e., \Msat$\ga 10^{10.8}$\Msun) within 0.5~Mpc; the large mass gap that we observe between the typical UMPEG and its most massive satellite implies that the \zs1.6 UMPEGs assembled through major mergers. Using observed satellite counts with merger timescales from the literature, we estimate the growth rate due to mergers with mass ratio of $\ge$~1:4  to be $\sim$13\%~Gyr$^{-1}$ (with a $\sim$$2 \times$ systematic uncertainty).  This relatively low growth rate is unlikely to significantly affect the shape of the massive end of the stellar mass function, whose evolution must instead be driven by the quenching of new cohorts of ultra-massive star-forming galaxies.  However, this growth rate is high enough that, if sustained to \zs0, the typical \zs1.6 \Mumpeg~$=10^{11.6}$\Msun\ UMPEG can grow into a  \Mstars~$\approx 10^{12}$\Msun\ brightest cluster galaxy (BCG) of a present-day massive galaxy cluster. Our observations favour a scenario in which our UMPEGs are main-branch progenitors of some of the present-day BCGs that have first assembled through major mergers at high redshifts and grown further through (likely minor) merging at later times.
\end{abstract}




\section{Introduction}

In the modern cosmic structure formation framework, dark matter halos form from initial density perturbations via gravitational collapse \citep{WhiteRees1978} and subsequently grow bottom-up via hierarchical merging \citep{WhiteFrenk1991, KauffmannWhite1993}.  Today, the most massive of these halos host galaxy clusters, at the centres of which reside Brightest Cluster Galaxies (BCGs) -- ultra-massive galaxies that are among the most extreme, most massive galaxies in the present-day Universe (stellar masses \Mstars $\ga 10^{11.5}$\Msun, and as high as $\sim$$10^{12}$\Msun; \citealt{Lidman2012, Liu2012}).   

Virtually all BCGs today are quiescent -- i.e., devoid of any significant star formation 
-- with very low specific star formation rates ($< \rm SSFR >$~$\sim0.001$~Gyr$^{-1}$;  \citealt{FraserMcKelvie2014}) that imply (current) mass build-up scales much longer than the present-day age of the Universe.  SSFRs in BCGs increase with increasing redshift, so that by \zs1.2 $<$SSFR$>$~$\sim0.1$~Gyr$^{-1}$ with very few, if any,  quiescent examples \citep{McDonald2016}.  
In agreement with this, models predict that BCG progenitors should be predominantly star-forming at high redshifts, $z>1$, but indicate that a small fraction may have been already massive and quiescent at these early times \citep{Contini2016}.  
Such Ultra-Massive Passively Evolving Galaxies (UMPEGs; \Mstars~$>10^{11.5}$\Msun), while very rare have
-- however --
been found at $z>1$ \citep[e.g.,][]{Marchesini2014, LARGE1}, with a handful of examples spectroscopically confirmed out to \zs4 \citep{Onodera2012, Belli2016, KadoFong2017, Stockmann2020, Forrest2020}.  If they formed as single objects through intense bursts of star formation, UMPEG progenitors must have had star formation rates (SFRs) of several hundred \Msun/yr \citep[e.g.,][]{LARGE1}.  Moreover, UMPEGs are very strongly clustered 
\citep{Cheema2020}, 
suggesting that they are associated with the high-$z$ progenitors of present-day massive galaxy clusters; UMPEGs could therefore be the direct, main-branch progenitors of some of the present-day BCGs.  Altogether, UMPEGs appear to be extreme examples among the high-$z$ galaxy population; because they are so extreme, they provide a unique opportunity to test our understanding of galaxy formation and evolution in the extreme regions of galactic parameter space that these monster galaxies occupy.   

Simulations \citep[e.g.,][]{DeLucia2007, RagoneFigueroa2018}  predict that a main-branch BCG progenitor (whether star-forming or quiescent) is often already well established by \zs1.5, and contains 10-30\% of the BCG's present-day stellar mass by that redshift.  Major mergers can be expected to play an important role in the formation of these ultra-massive galaxies (UMGs) because of dynamical friction \citep{Ostriker1975}, whose effect increases with increasing satellite mass \citep{Chandrasekhar1943}  As a result, massive galaxies at the core of their dark matter halo can be expected to merge relatively quickly to form an even more luminous and massive galaxy, leaving as evidence of such merging a substantial luminosity gap between the brightest and second-brightest galaxy in the system \citep{Ostriker1975, Jones2000, Tal2012}.  The more detailed semi-analytic formation models \citep[e.g.,][]{DeLucia2007, Contini2016} indeed predict a significant major-merger phase, expected to occur at high redshift ($z>1$), although in-situ star formation (before quenching occurs) and minor mergers can also contribute to the growth.

At low and intermediate ($z\la1$) redshifts, observations suggest that mergers continue to add mass to the BCGs \citep[e.g.,][]{Edwards2012, Liu2012, Liu2015, Burke2013, Lidman2013}. However, to test the merger  paradigm at high redshifts, $z\ga1$,  requires first the ability to identify high-$z$ BCG progenitors. Such progenitor identification is non-trivial given that -- as mentioned earlier -- the progenitors are expected to evolve substantially through mergers (as required by the model).

One way to proceed is to modify the abundance-matching technique by correcting for the effects of merging using predictions from numerical simulations \citep[e.g.,][]{Marchesini2014, Vulcani2016, Cooke2019}.  These studies report the detection of significant number of UMG progenitors at high redshift, $z\ga1$ and find that while many of them are star-forming, a non-negligible fraction is already quenched. At $1\la z \la 2$, growth in UMG stellar masses comes from in-situ star formation, which is estimated directly from observational data, and from mergers, whose effect is estimated from simulations \citep{Cooke2019}.  While supporting the importance of mergers at high redshift, this approach does not test it directly as it uses hierarchical halo merger models to both help identify BCG progenitors, and to estimate merger and mass growth rates \citep{Cooke2019}.  Merger rates can be instead be more directly estimated from the data by counting the incidence of 
close companions or disturbed morphologies, as done by \citep{Vulcani2016} and \citep{Zhao2017}, who find evidence for mergers at $z>1$, although still relying on abundance matching for progenitor identification.
 
An alternative approach to testing the merger scenario is to identify BCG progenitors in a way that is less dependent on assumptions inherent in 
galaxy 
abundance matching, whether corrected or not using hierarchical merger trees. In \citep{LARGE1} we used wide-area imaging to identify a population of $z\sim1.6$ UMPEGs, 
and their extremely strong clustering ($r_0 = 29.77 \pm 2.75 \ h^{-1}$Mpc) suggests that they are associated with halos of mass $M_{\rm halo} \sim 10^{14.1} h^{-1}$\Msun.  Under the assumption of smooth halo mass growth, these \zs1.6 halos are likely to be the progenitors of present-day massive ($M_{\rm halo}\sim 10^{15}$\Msun) galaxy clusters
\citep{Cheema2020}.  
While only approximately one out of eight of such massive \zs1.6 halos appear to contain an UMPEG (the remaining halos presumably contain star-forming UMGs, or collections of lower-mass pre-merger components), the very high stellar masses of our UMPEGs and their very strong clustering suggest that they are direct, main-branch progenitors of some of the present-day BCGs in massive clusters. These \zs1.6 objects thus provide a sample of plausible proto-BCGs with which to test the BCG hierarchical growth scenario at high redshift. 

In this paper we will examine the environments of these \zs1.6 BCG progenitors.  The depth of our dataset is not sufficient over most of its area to find normal, sub-$M^*$, galaxies around them, and consequently we are not in a position to see directly if they reside at centres of (proto)clusters. Instead, we will search for evidence of major mergers in the formation and growth of our UMPEGs.  We will do so not by searching for merger signatures such as disturbed morphologies or tidal features \citep[e.g.,][]{Patton2000, Lotz2008, Bridge2010} as our data are also too shallow to detect such low-surface-brightness features at \zs1.6. Instead, we will search for (relatively) massive companions that represent a reservoir of material for future mergers. The number of such massive companions, combined with dynamical friction timescales, will give us an estimate of the mass growth due to mergers. At the same time, the presence of a significant mass (or luminosity) gap between the UMPEGs and their most massive (or luminous) satellites will provide evidence that major mergers have already happened. 
 
Throughout this work we use the AB magnitude system \citep{Oke1974} and  assume the \citet{Chabrier2003} stellar initial mass function (IMF) when calculating stellar masses of galaxies. We adopt the flat $\Lambda$ cosmology with  $\Omega_{m,0}=0.3$,  $\Omega_{\Lambda,0}=0.7$, and Hubble constant of $H_{0} = 70$ km s$^{-1}$Mpc$^{-1}$. In this cosmology, 1~Mpc~(physical) corresponds to 118.0\arcsec, or $\sim$~2\arcmin, at \zs1.6 \citep{Wright2006}.

\section{Data and sample selection}\label{sec:data}

The detailed description of our catalogs and sample selection is given in \cite{LARGE1} so here we only summarize the key details.   

We selected \zs1.6 galaxies using an adaptation of the \cite{Daddi2004} $BzK$ technique developed by \cite{ArcilaOsejo2013}.  This ``\gzK\ technique'' selects \zs1.6 star-forming (\SFgzK) and passively-evolving (\PEgzK) galaxies using a combination of optical and near-IR photometry.  For the optical data we used the  $g$ and $z$ images from the T0006 release of the CFHT Legacy Survey (CFHTLS, \citealt{CFHTLS-T0006}), specifically all four of its Deep fields (D1, D2, D3, and D4), and two of the Wide fields (W1 and W4).  For the NIR data in W1 and W4 we used the $K_s$ images from the Visible Multi-Object Spectrograph (VIMOS) Public Extragalactic $K_{s}$ Survey Multi-Lambda Survey (VIPERS-MLS; \citealt{refIda}); meanwhile, in the Deep fields we used the \Ks\ and $H$-band data from the T0002 release of the WIRCam Deep Survey (WIRDS;  \citealt{refIdb}).

Requiring overlap between the optical and NIR datasets, and after masking areas around bright stars, low-SNR regions, and other artifacts, our data cover 25.09 deg$^2$ in the two Wide fields and 2.51 deg$^2$ in the Deep fields, giving a total area of 27.6 deg$^2$.  In the Wide fields we reach 90\% detection completeness at \Ks=20.5 AB;  in the Deep D1, D3, and D4 fields we reach 50\% detection completeness at \Ks=23.5 AB, while in D2 (the COSMOS field) we reach it at \Ks=23.0 AB.  

We used {\sc SExtractor} \citep{refId10} to perform object detection in the \Ks-band and then matched-aperture photometry in other bands. Applying the \gzK\ selection technique of \cite{ArcilaOsejo2013} to these data gave us 8,756 \gzK\ galaxies with 19.25$<$\Ks$<$20.25 in the Wide fields and 53,988--65,586 (the lower number is for secure \zs1.6 galaxies and the upper number includes some possible lower-$z$ interlopers) with 19.25$<$\Ks$<$23.5 in the Deep fields.   Comparison of our sample in the COSMOS (CFHTLS D2) field with the catalog of \cite{Muzzin2013}, shows that \gzK\ galaxy redshift distributions very with type (\PEgzK\ and \SFgzK) and magnitude: the peak is at \zs1.5 for the brightest galaxies and shifts to \zs1.7 for fainter galaxies (\Ks~$\sim$22--23), while the FWHM of the distribution is $\sim$0.8 for \PEgzK\ galaxies and $\sim$1.1 for \SFgzK\ (see Fig.~6 of \citealt{LARGE1} for an illustration). 

The brightest among the \PEgzK\ galaxies are of special interest.  In order to be already quiescent at \zs1.6 and yet very bright in the NIR suggests they must be very massive and thus must have formed under extreme conditions at even higher redshifts.  We thus defined in \cite{LARGE1} a sample of Ultra-Massive Passively Evolving Galaxies (UMPEGs) to be \PEgzK\ galaxies with  \Ks$<$19.5. This selection corresponds to quiescent galaxies with stellar masses \Mstars$\ga$$10^{11.5}$\Msun, and with mean stellar mass  $<$\Mstars$>$ = $10^{11.6}M_{\odot}$. Our sample of UMPEGs consists of 61 objects, the vast majority (all but six) located in the two Wide fields (W1 and W4).  For full details of the object detection, photometry,  UMPEG selection, 
and stellar mass estimation,
please see \cite{LARGE1}.

\section{Satellite counts}\label{sec:satelliteCounts}

\begin{figure}
\begin{centering}
\includegraphics[scale=0.8]{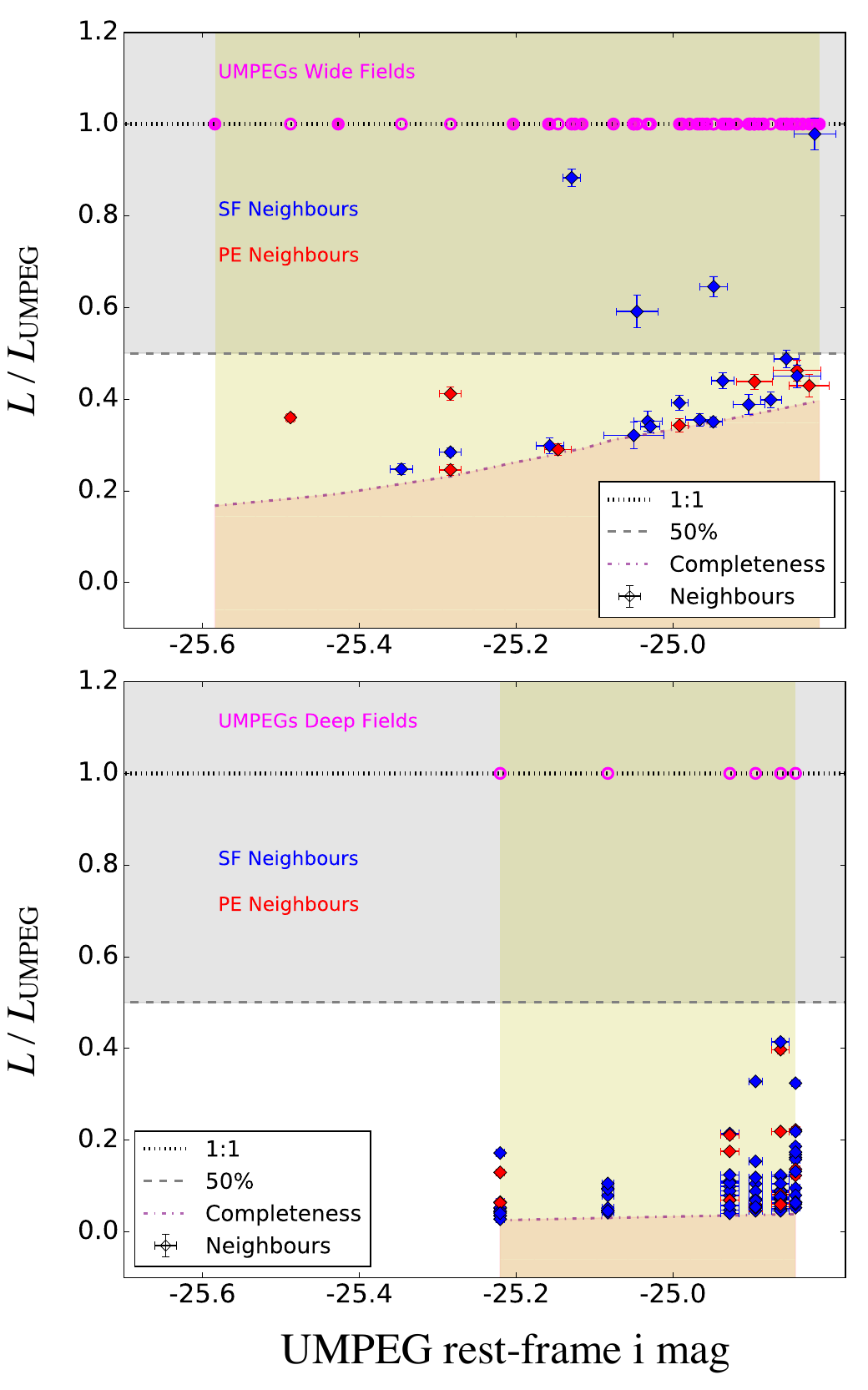}
\par\end{centering}
\protect\caption{Satellite galaxy candidates.  Luminosity ratios (satellite candidate compared to that of its UMPEG) are plotted as function of UMPEG luminosity in rest-frame $i$-band.  The top panel shows data in the Wide fields, and the bottom panel is for the Deep.  The yellow area shows  the magnitude range of our UMPEGs;   open magenta symbols identify UMPEGs with satellite candidates, whereas   filled magenta points are for UMPEGs with no satellites.  The gray shaded area represents the positions these companions would occupy if they were at least 50\% as bright as the central UMPEG. The completeness limit for each sample, 
reported in Sec.~\ref{sec:data},
is shown with a dot-dashed line. }
\label{fig:companions}
\end{figure}

\begin{figure}
\begin{centering}
\includegraphics[scale=0.55]{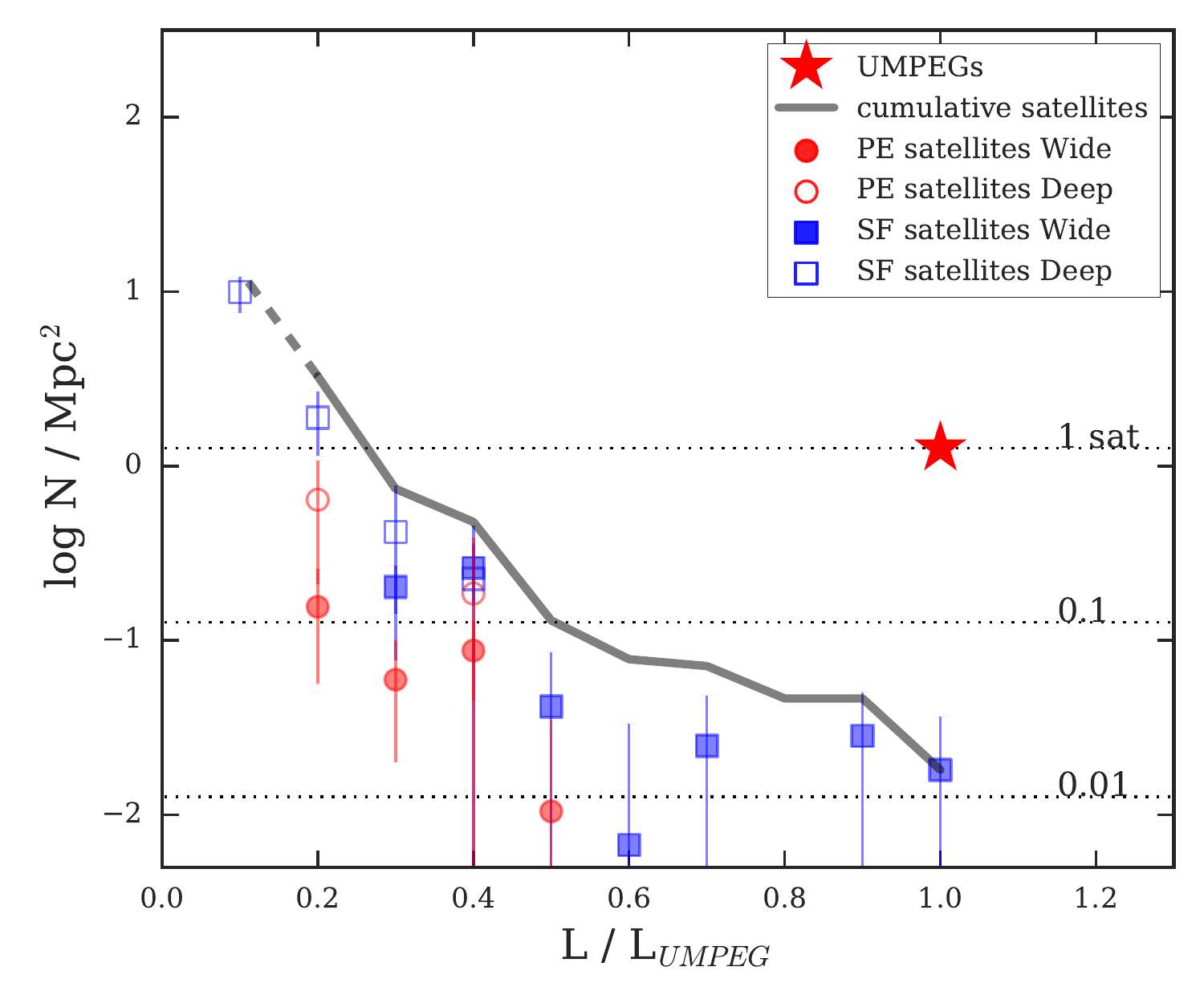}
\par\end{centering}
\protect\caption[Surface area density of companion gzK$_{s}$ galaxies around UMPEGs.]{
The background-subtracted surface number density of companions within 0.5~Mpc of UMPEGs as a function of rest-frame $i$-band luminosity ratio.   The three dotted horizontal lines show (top to bottom)  1, 0.1, and 0.01 satellites per UMPEG.   The black curve shows the cumulative number density of satellites: solid where we are complete and dashed to indicate the lower limit where we do not detect \PEgzK\ galaxies. It is clear that UMPEGs lack companions of comparable luminosity even in this figure, which does not remove unassociated fore/background projections.  Even at luminosities $\sim 0.2 \times L_{\text{UMPEG}}$ there is only one companion (physical + projected) per UMPEG. 
\label{fig:N-vs-Lratio}}
\end{figure}


Halos of mass $\sim10^{13}$\Msun\ are expected to have virial radii of 0.5 Mpc  at \zs1.6 (\citealt{Munoz-Cuartas2011}).   Our clustering analysis suggests that our \zs1.6 UMPEGs reside in halos that are an order of magnitude more massive \citep{Cheema2020}, so searching for potential satellite galaxies within a 0.5~Mpc radius 
seems justified as it is likely to yield a physically-associated population.  We perform this search by conducting a census of all \gzK\ galaxies located in 0.5~Mpc (projected) proximity to each UMPEG followed by  a statistical correction to account for the contamination due to chance projections. 

Figure~\ref{fig:companions} shows our census of the potential satellites of our UMPEGs in the Wide (top panel) and Deep (bottom panel) fields. The rest-frame $i$-band luminosities of the UMPEGs, derived from their \Ks\ magnitudes, are shown on the horizontal axes, while the vertical axes show the $i$-band luminosity ratios ($L_{\text{comp}}/L_{\text{UMPEG}}$) for the (potential) satellites.  The candidate satellites in Fig.~\ref{fig:companions} include physically unassociated \gzK\ fore/background galaxies, so the number of objects shown can be regarded as an upper limit on the number of physically-associated companions.  However, even with this simple analysis it is clear that UMPEGs have extremely few luminous companions:  of the 61 UMPEGs in our sample, only four have a companion with luminosity ratio $L_{\text{comp}}/L_{\text{UMPEG}} > 0.5$ (objects above the dashed line).  Only at $L_{\rm comp}/L_{\rm UMPEG} < 0.2$ do we start seeing significant numbers of companions, although many of these are likely to be physically-unassociated chance projections. 

Of course, we currently have no way of individually identifying which of the objects in Fig.~\ref{fig:companions} are physically associated with our UMPEGs and which are chance projections.  We can, however, statistically correct for this foreground/background contamination using the well established approach recently employed by a number of studies that characterize the satellite galaxy populations around massive galaxies \citep[e.g.,][L.~Chen et al., submitted to MNRAS]{Tal2012, Tal2013, Kawinwanichakij2014, Hartley2015}.   In this approach, the (fore/background-contaminated) number counts of galaxies within a projected cylinder centred on the central galaxy are corrected by subtracting the magnitude-dependent number counts inferred from similar cylinders placed at random positions in the survey area.  In our case, to compute the background we randomly place within our survey area three hundred apertures with radius 0.5~Mpc each (ensuring that they do not overlap with the 0.5~Mpc apertures centred on our UMPEGs), and then count \gzK\ galaxies within them to arrive at the background number density as function of magnitude and galaxy type (\PEgzK\ or \SFgzK). Subtracting these numbers from the counts of \gzK\ galaxies around the UMPEGs statistically removes physically-unassociated, projected foreground/background \gzK\ galaxies:  this correction statistically deems most UMPEG companions in the Wide fields as unphysical, and $\sim$60\% in the Deep fields.  Applying this correction gives us the background-corrected observed surface density of true, physically-associated satellites, $\Sigma_{\rm obs}$.

We next need to correct for an additional source of incompleteness which affects the probability of detecting satellites.  Here we note that the redshift distribution of UMPEGs is approximately Gaussian-shaped \citep{LARGE1} and that this is a reflection of the \gzK\ selection window rather than a real physical decrease in the abundance of ultra-massive galaxies away from \zs1.6.  This means that objects whose redshifts are away from the peak of the redshift distribution have a reduced probability of detection.  This effect affects the detectability of UMPEGs, but because we do not analyze the environments of such undetected UMPEGs, our satellite statistics are not affected.  However, the detectability of satellites UMPEGs that {\emph{are}} detected but are at redshifts offset from the peak of the redshift distribution is also reduced, and we need to account for these undetected satellites. We correct for this incompleteness by statistically weighing $\Sigma_{\rm obs}$ as 
\begin{equation}
\Sigma_{\rm true}=\Sigma_{\rm obs}\times\frac{\int N_{\mathrm{UMPEG}}(z)dz}{\int N_{\mathrm{UMPEG}}(z)\times N_{\rm comp}(z)dz},
\label{eq:fcorrection-1-1}
\end{equation}
where $\Sigma_{\rm true}$ represents the true surface density of satellites, corrected for both fore/background galaxies and satellite detection completeness. The $N(z)$ are the redshift probability distributions of UMPEGs or companions. For these redshift distributions, we adopt the photometric redshift distributions shown in Fig.~6 of \cite{LARGE1}, noting that the probability density distribution $N(z)$  is different for UMPEGs and companions given their different $K_s$ magnitudes and the fact that companions can be star-forming or passive.  In Eq.~\ref{eq:fcorrection-1-1} the denominator accounts for the fact that companions will only be detected when both their central UMPEGs {\it and} they themselves are detected, and this is proportional to both the UMPEG and companion detection probabilities;  the numerator accounts for the redshift-dependent detection probability of the UMPEGs. 

Figure~\ref{fig:N-vs-Lratio} shows the results of applying this procedure, binned in bins of luminosity ratio and separated for \SFgzK\ and \PEgzK satellites.  The gray curve shows the cumulative number density of satellites (of both types) summed starting with the most luminous. Note that in this Figure we have converted the observed quantities (number per arcmin$^2$) into physical ones (number per Mpc$^2$) using our adopted cosmology.  

It is clear in Fig.~\ref{fig:N-vs-Lratio} that  UMPEGs have virtually no physically-associated satellites with $L/L_{\text{UMPEG}} \ga 0.5$, and even summing down to  $L/L_{\text{UMPEG}} \sim 0.25$ there is only on average one satellite per UMPEG. The shape of the distribution, with its low number of luminous satellites followed by an increase in satellite number at lower masses, is reminiscent of the  "gap" reported by \cite{Tal2012} in the luminosity function of \zs0.3--0.7 Distant Red Galaxies (DRGs). The details of their distributions are not directly comparable to those of our's because we normalized our satellite UMPEG luminosities, whereas \cite{Tal2012} left heir DRG luminosities un-normalized. Nevertheless, the size of the luminosity gap -- which \citeauthor{Tal2012} defined as the point at which the cumulative satellite distribution reaches one satellite per central -- is similar:  in their DRG sample they found it to be $L_{\rm sat}/L_{\rm DRG}\sim0.3$, which is similar to the $L_{\rm sat}/L_{\rm UMPEG} \sim0.25$ that we find for our UMPEGs. The implications for our \zs1.6 UMPEGs are similar to those for the intermediate-redshift DRGs of \citeauthor{Tal2012}:  the paucity of luminous companions suggests that our UMPEGs may have formed a significant time before the redshift of observation -- \zs1.6 in our case -- and that their subsequent growth will be moderate and mainly through minor mergers.  We will expand on these points in Sec.~\ref{sec:discussion}. 

\section{Discussion}\label{sec:discussion}

\begin{figure}
\begin{center}
\includegraphics[width=0.51\textwidth]{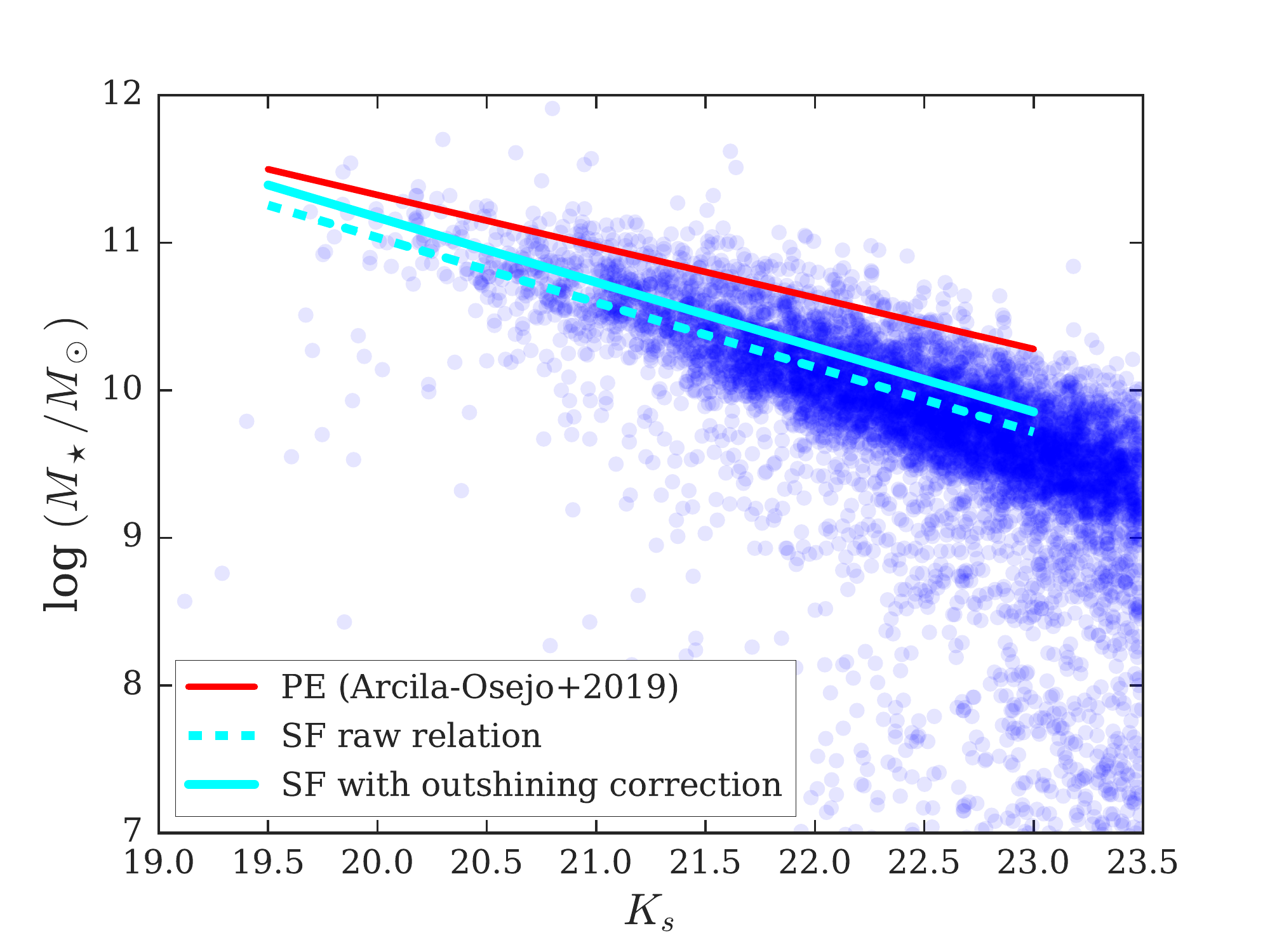}
\end{center}
\caption[]{ 
The mass-magnitude relation for star-forming \gzK\ galaxies in the COSMOS field.  The dashed blue line is a fit to the data as they are, while the solid blue line represents that raw relation adjusted for the effects of outshining in stellar mass estimates following the empirical prescrtiption of \cite{Sorba2018}.  The red line shows the mass-magnitude relation for quiescent, PE-\gzK\ galaxies from \cite{LARGE1}.
}
\label{fig:SF-mag-to-mass}
\end{figure}

\subsection{Masses of UMPEG satellites}\label{sec:masses}

UMPEGs are already extremely massive (\Mumpeg $> 10^{11.5}$\Msun), but it is interesting to consider how much they could grow by absorbing their satellites as these lose momentum via dynamical friction and in-spiral towards the centre of the host dark matter halo. To do so, we first need to estimate the stellar masses of the satellites before we estimate how quickly these satellites will merge with their UMPEGs. 

In \citet{LARGE1} we determined an empirical  \Ks-\Mstars\ relation for quiescent galaxies by matching our \Ks\ magnitudes of \PEgzK\ galaxies in the COSMOS field with their stellar masses from the \citet{Muzzin2013} catalog that's based on SED fitting of multi-band photometry. Fitting the resulting distribution gave us the relation
\begin{equation}
\label{eq:PE-mag-to-mass}
\log [M_{\star}^{\text {PEG}}/M_\odot] = - 0.348 K_s + 18.284,
\end{equation}
which we can apply to the quiescent satellites. 
Because many of the satellites are star-forming, we also need a similar relation for \SFgzK\ and so we now repeat the above matching but for \SFgzK\ galaxies and show the results in Fig.~\ref{fig:SF-mag-to-mass}, where the magnitudes are our total \Ks\ magnitudes, the stellar masses are again from the catalog of \cite{Muzzin2013}, and the dashed blue line shows the line of best fit to these  data.  Note, however, that there is growing evidence that spatially unresolved spectral energy distribution (SED) fitting (such as that in \citealt{Muzzin2013} but also in the vast majority of other studies) underestimates stellar masses of star-forming galaxies (\citealt{Sorba2018}; see also \citealt{Zibetti2009}, \citealt{Sorba2015}, \citealt{Abdurrouf2018}). The effect arises because old stars, which can contain the bulk of a galaxy's stellar mass, are masked in broadband photometry that's used in SED fitting by much brighter but less numerous young stars (\citealt{Sorba2015, Sorba2018}; see also \citealt{Sawicki1998}). As shown in \cite{Sorba2015, Sorba2018}, this systematic mass under-estimate depends on specific star formation rate (SSFR) and so we correct the raw best-fit relation (dashed line in Fig.~\ref{fig:SF-mag-to-mass}) using the prescription from \citet[][their Eq.~6]{Sorba2018} taking $\text{SSFR}=10^{-8.8}$~$\text{yr}^{-1}$, as appropriate for high-mass  (\Mstars~$\sim 10^{10}$~\Msun) star-forming galaxies at \zs1.6 \citep{Whitaker2014, Johnston2015}.  The result is an increase of  star-forming galaxy stellar masses by $\Delta M_\star^{\rm SFG}$ = 37\% on average, as compared to the masses from spatially unresolved SED fits. The resulting mass-magnitude relation, corrected for the outshining effect, is then given by
\begin{equation}
\label{eq:SF-mag-to-mass}
\log [M_{\star}^{\text{SFG}}/M_\odot] = - 0.439 K_s + 19.958,
\end{equation}
and is shown with the solid cyan line in Fig.~\ref{fig:SF-mag-to-mass}.  The relation for quiescent galaxies (Eq.~\ref{eq:PE-mag-to-mass}) is shown with the red line for reference.

\begin{figure}
\begin{center}
\includegraphics[scale=0.55]{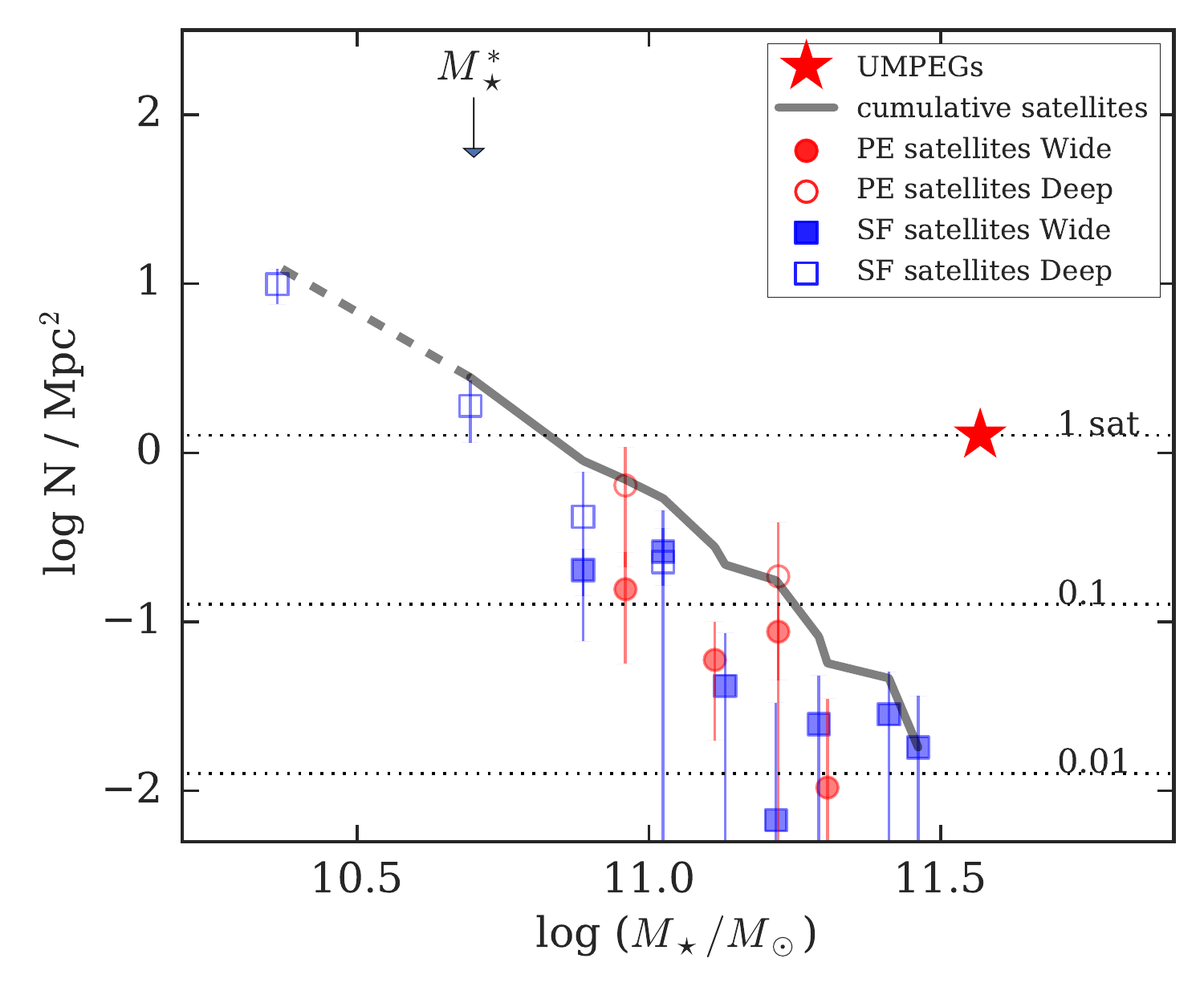}
\end{center}
\caption[]{ 
The background-subtracted surface number density of satellites within 0.5~Mpc of UMPEGs as a function of stellar mass. The black curve shows the cumulative number density of satellites: solid where we are complete and dashed to indicate the lower limit where we do not detect \PEgzK\ galaxies. The three dotted horizontal lines show (top to bottom) 1, 0.1, and 0.01 satellites per UMPEG. The location of the characteristic Schechter mass, $M_\star^*$, is indicated. 
}
\label{fig:Ncomp-vs-logM}
\end{figure}

Using the \Ks-\Mstars\ relations for quiescent (Eq.~\ref{eq:PE-mag-to-mass}) and star-forming (Eq.~\ref{eq:SF-mag-to-mass}) galaxies, and assuming that UMPEGs have a mass of \Mumpeg$=10^{11.57}$\Msun\ (the mean mass of our UMPEG sample), we recast the data of Fig.~\ref{fig:N-vs-Lratio} in terms of stellar masses.  The result is plotted in Fig.~\ref{fig:Ncomp-vs-logM}. As that Figure shows, on average, UMPEGs have virtually no massive  physical companion galaxies (mass ratios \Msat\ : \Mumpeg~$\ga$ 1:3).  One has to integrate down to \Msat\ : \Mumpeg~$\approx$ 1:5.5 to find one satellite per UMPEG, on average. 

We note that for a typical UMPEG with \Mumpeg~$\approx 10^{11.6}$\Msun, satellites of 1:5.5 mass ratio have stellar masses \Msat = $0.18\times$\Mumpeg$\approx 10^{10.8}$, which is similar to the characteristic Schechter mass of $M^{*}_{\star} \sim 10^{10.6-10.8}$ for quiescent \zs1.6 galaxies  \citep[indicated with the arrow in Fig.~\ref{fig:Ncomp-vs-logM}; ][]{ArcilaOsejo2013, Ilbert2013, Muzzin2013b, Tomczak2014, LARGE1}.  Consequently, our ``low-mass'' UMPEG satellites are in fact very massive themselves.  Nevertheless, in relative terms, they are of significantly lower mass than the mass of the UMPEGs that they are associated with and with which they will likely eventually merge. 

We will consider how mergers -- both minor and major -- contribute to the growth of UMPEGs in the next section. 

\subsection{UMPEG growth through mergers}

\subsubsection{Growth rate estimate}\label{sec:growthRate}

\begin{figure}
\begin{center}
\includegraphics[scale=0.55]{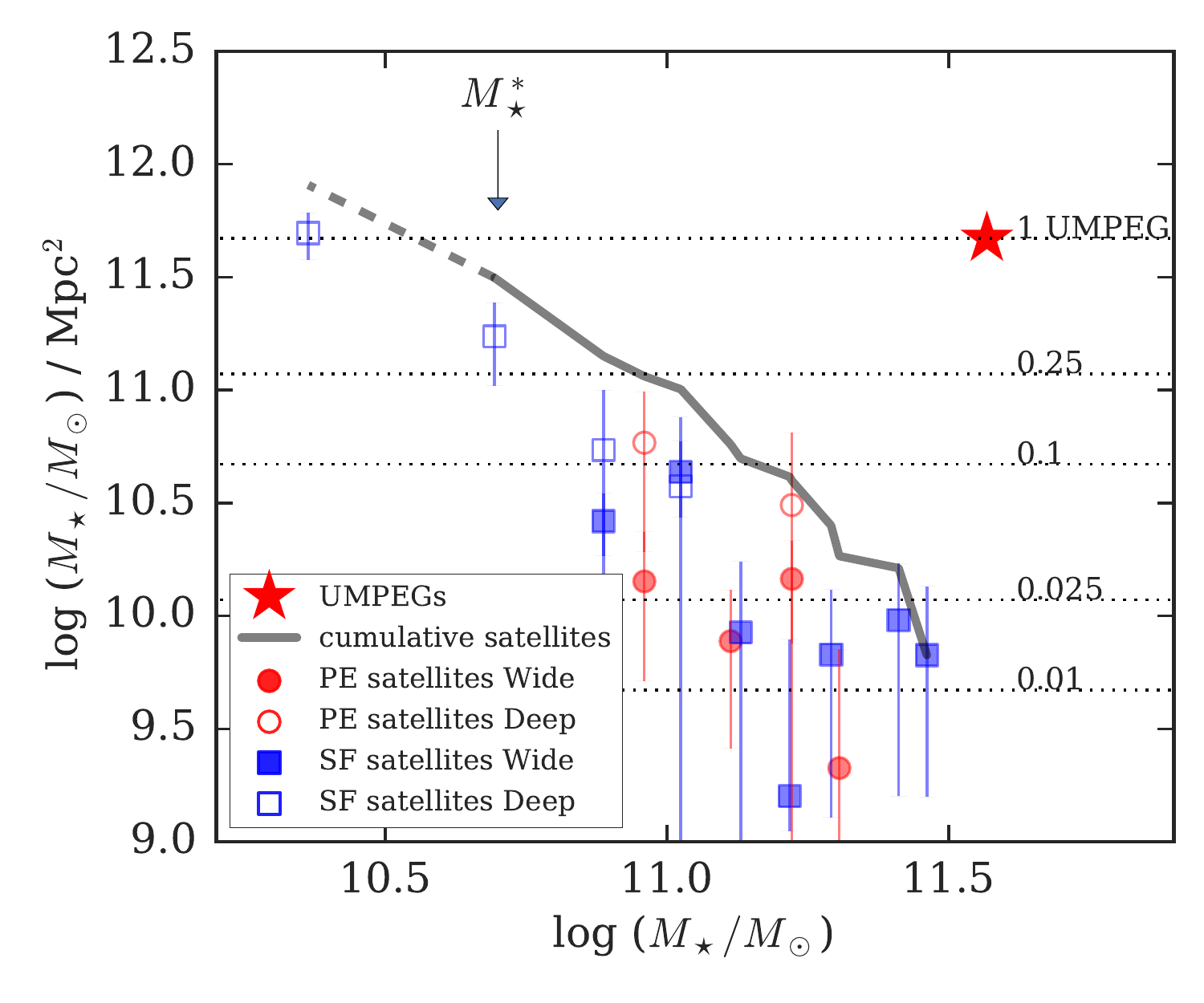}
\end{center}
\caption[]{ 
The background-subtracted surface mass density of satellites within 0.5~Mpc of UMPEGs as a function of stellar mass. The black curve shows the cumulative number density of satellites: solid where we are complete and dashed to indicate the lower limit where we do not detect \PEgzK\ galaxies. The location of the characteristic Schechter mass, $M_\star^*$, is indicated. The  dotted horizontal lines show (top to bottom) 1, 0.25, 0.1, 0.025, and 0.01 times the mass density of a typical (\Mstars=10$^{11.57}$\Msun) UMPEG if that mass were distributed over the same projected area as the mass contained in the satellites.}
\label{fig:SMD-vs-logM}
\end{figure}

It is interesting to consider how much (or how little) UMPEG masses can grow by absorbing stellar material from the galaxies we observe in their environment.  For this, consider Fig.~\ref{fig:SMD-vs-logM} which shows the stellar mass surface density in satellites out to our 0.5~Mpc search radius. The black line shows the cumulative surface stellar mass density contributed by the satellites and it is clear that the sum of the masses of all major satellites of UMPEGs is quite small:  for example, down to \Msat~ $=10^{11.0}$\Msun~$\approx0.25\times$\Mumpeg, the stellar masses of all the satellites add up just under a quarter the mass of the central UMPEG; down to \Msat$=M^*_\star = 10^{10.7}$\Msun, the mass contained in satellites is only $\sim$40\% times the mass of the UMPEG.  Clearly, there is not a lot of stellar material available for growth via major (\Mstars:\Mumpeg $>$ 1:4) mergers in the vicinity of the UMPEGs.  

We can make a rough estimate of the mass accretion rate, for which we need a merger timescale.  First, we make a crude estimate of the merger timescale, $T_{\rm merge}$, using the dynamical friction formula from \citet[][their Eq.~8]{Jiang2008}.  We assume a value of $\varepsilon=0.5$, which is independent of the masses of the interacting galaxies (\citealt{Jiang2008}), and circular velocity $v\sim$860 km/s (a velocity that relates to a high likelihood of interaction for close pairs; \citealt{Patton2000}). We also estimate $T_{\rm merge}$ using the formula from \citet[][their equation for $\Delta v <  3000$ km~s$^{-1}$]{Kitzbichler2008} that is calibrated for close pairs but does provide an estimate of $T_{\rm merge}$ alternative to that of \cite{Jiang2008}.

Applying the two $T_{\rm merge}$ estimators to the UMPEG companions and summing over the sample after appropriately weighting by the fore/background counts, allows us to estimate the merger timescales, and then -- incorporating our satellite mass estimates -- to arrive at the mass accretion rates. The two merger timescale prescriptions yield accretion rates that are consistent for high-mass satellites but differ by up to a factor of $\sim2$ at lower \Msat.   Taking the average of the two gives a growth rate of $\sim$13\% Gyr$^{-1}$ for an UMPEG with \Mumpeg~$ = 10^{11.6}$\Msun, with an uncertainty of a factor-of-two that's driven by the systematic uncertainty in the $T_{\rm merge}$ estimate.

\subsubsection{Growth since \zs1.6}\label{sec:growthAfter}

While we do not know whether the accretion rate remains constant for UMPEGs, there are indications for lower-mass central galaxies that satellites that are consumed are almost exactly replenished by new arrivals from beyond the halo \citep{Tal2013, Hartley2015}.  It thus seems plausible that newly arriving satellites keep the UMPEG halos stocked with new arrivals as existing satellites sink towards the halo's centre due to dynamical friction and are consumed by merging with the UMPEGs. Taking the estimated growth rate of $\sim$13\% Gyr$^{-1}$ at face value would grow our fiducial UMPEG from \Mumpeg=$10^{11.6}$\Msun\ at \zs1.6 to  $M_\star^{z=0}=10^{11.9-12.1}$\Msun\ by \zs0 (the mass range reflects linear or exponential growth in the accretion rate. Notably, this forecast \zs0 UMPEG mass is comparable to the masses of BCGs in present-day massive galaxy clusters  (\Mstars $\sim 10^{11.7}$\Msun with a few examples at \Mstars $> 10^{12}$\Msun; see compilation by \citealt{Lidman2012}). For example, the mass of NGC 4874 (one of the central galaxies of the Coma cluster) is  \Mstars$\sim10^{12.0}$ \citep{Veale2017}.

Our mass growth calculation is crude as it assumes a constant merger rate (whereas -- at least for lower-mass galaxies -- merger rates   decrease with time; e.g.,  \citealt{Patton2002, Bridge2010}) and, moreover,  does not account for accretion of low-mass satellites or for possible re-ignition of star formation in the UMPEGs.  However,   this rough calculation  does yield a projected \zs0 UMPEG mass that is consistent with the masses of massive cluster BCGs at \zs0 and thereby with the idea, first suggested on the basis of clustering measurements in \citet{Cheema2020}, that our \zs1.6 UMPEGs are the direct progenitors of some of the central BCGs of massive present-day galaxy clusters.
 
Our mass growth rate estimate appears to be somewhat higher than that reported by \citet{Vulcani2016}, who used the modified galaxy abundance-matching method of \citet{Marchesini2014} to select the progenitors of massive (\Mstars $> 10^{11.8}$ \Msun) \zs0 galaxies at earlier epochs and then used counts of their satellites to infer merger mass growth.  \citeauthor{Vulcani2016} report that a present-day \Mstars~$\ga10^{11.8}$ massive galaxy progenitor, which in their abundance-matching selection has a typical \zs1.7 mass of \Mstars$\sim$10$^{11.35}$, can only reach its final mass through a combination of mergers and in-situ star formation (roughly half of their progenitors are star-forming at $1.5\ga z\ga 2$). This would appear at odds with our conclusion, but we note some important differences between the two  studies: First, our UMPEGs are selected via clusering, rather than through galaxy abundance matching, so they may represent a different population; this, and the fact that our UMPEGs are selected to be quiescent -- while the \citeauthor{Vulcani2016} objects include a large number of star-forming systems -- suggests that UMPEGs could be preferentially older systems, likely located in regions that collapsed earlier in the history of the Universe than those that host the typical object in the \citeauthor{Vulcani2016} sample. Second, our UMPEGs are more massive than the \citeauthor{Vulcani2016} objects, which means they require less mass growth to reach the same final \zs0 mass. And third, it is likely that the masses of the star-forming (but not quiescent) galaxies in the \citeauthor{Vulcani2016} sample are underestimated \citep[][and Sec.~\ref{sec:masses}]{Sorba2018};  accounting for this underestimate would make more satellite mass available for merging, making growth through mergers more efficient than estimated by \citeauthor{Vulcani2016}; furthermore, the mass bias would also make the star-forming centrals in the \citeauthor{Vulcani2016} more massive, making it even easier for them to reach the target \zs0 mass. For these reasons, and also because our mass growth estimates are rather crude (as discussed earlier), we do not feel that the results of the two studies are in conflict. On the contrary, given the systematic effects discussed above, the results may give a consistent picture in which the \zs1.6 quiescent progenitors of present-day massive centrals grow through mergers alone, while their star-forming progenitors grow through a combination of mergers and (until quenched) in-situ star formation.

As a final point in this section we briefly discuss the effect of mergers on the evolution of the massive end of the stellar mass function.  The relatively low mass growth rate from mergers that we estimate suggests that the evolution of the massive end of the stellar mass function of quiescent galaxies is not strongly affected by mergers.  Instead, the growth seen from \zs1.6 to intermediate and low redshifts \citep{ArcilaOsejo2013, Muzzin2013b, Ilbert2013, Moutard2016b, LARGE1} must be primarily driven by the arrival of newly-quenched massive galaxies joining the quiescent population. 

\subsubsection{Growth before \zs1.6}\label{sec:growthBefore}

It is also interesting to consider  how our \zs1.6 UMPEGs may have grown from higher-redshift progenitors, and a clue lies in the presence of the gap -- noted in Sec.~\ref{sec:satelliteCounts} -- between the mass (or luminosity) of the UMPEG and the mass (or luminosity) of its most massive (luminous) companion.  In terms of luminosity, the gap is $L_{\rm 1st~sat}/L_{\rm UMPEG} \sim 0.25$ (i.e., $\sim 1.5$ mag), and is of similar size to the luminosity gap found for satellites of massive quiescent DRGs at $z=0.3-0.7$ by \cite[][$L_{\rm 1st~sat}/L_{\rm DRG} \sim 0.3$ ]{Tal2012}.  For our \zs1.6 UMPEGs, the  gap in mass is  $M_{\star}^{\rm 1st~sat}/M_\star^{\rm UMPEG} \sim 5.5$ (Fig.~\ref{fig:Ncomp-vs-logM}).

The presence of a significant luminosity gap between the first and second most luminous group members can be interpreted as indicative of the age of the hosting environment \citep[e.g.,][]{TremaineRichstone1977, VanDenBosch2007, Milosavljevic2006, Raouf2018,Dariush2010}.  This is because the most massive group members, which are the ones most strongly affected by dynamical friction, can be expected to merge rapidly, leaving behind only the less-massive satellites as these suffer less deceleration due to dynamical friction.  Major mergers thus provide a natural explanation for the presence of the luminosity (and mass) gap in our UMPEG environments.

Further evidence for this scenario is provided by the fact that about 10 percent of our original UMPEG candidates, excluded from the sample, appear to be double-cored systems (see \citealt{LARGE1}). These systems may be late-stage mergers of pairs of massive and already-quiescent UMPEG building blocks (so-called ``dry mergers'').  Of course, mergers that build UMPEGs might also be wet (i.e., gas-rich and resulting in additional star-formation), but we have no immediate way of observationally linking our UMPEGs to such wet merger progenitors with the present data. We note that our UMPEGs cluster extremely strongly (see 
\citealt{Cheema2020})
-- much more strongly that do potential high-$z$ massive wet merger candidates such as  sub-millimetre galaxies (SMGs, e.g., \citealt{Blain2004, Hickox2012, Wilkinson2017}) or Dust Obscured Galaxies (DOGs, e.g., \citealt{Brodwin2008, Toba2017}). It therefore seems that UMPEGs are associated with more massive, rarer dark matter halos than
 is 
 typical of high-$z$ starbursts.  This clustering argument does not rule out the possibility of wet mergers as UMPEG progenitors, but it does suggest that if SMGs and DOGs are wet mergers, the majority of them are not the wet mergers that later quenched to become UMPEGs. 

It is also interesting to consider if the mass accretion rate we estimated in Sec.~\ref{sec:growthAfter} is consistent with the major merger scenario.  By construction, this merger rate estimate is for major
mergers
 ($M$:\Mumpeg = 1:4). Applying the $\sim$13\% Gyr$^{-1}$ estimated growth rate to our fiducial \Mumpeg = $10^{11.57}$\Msun\ galaxy, and assuming for simplicity that major mergers are the only mass growth mechanism, we get a $z=4$ mass of \Mstars  = $10^{11.43}$\Msun\ (given the short time-span between $z=4$ and $z=1.6$, the numbers are almost the same for  linear and exponential growth and so we took the average of the two here). This growth could, for example, be accomplished through a single $\sim$3:1 merger ($10^{11.43}$\Msun\ + $10^{11.00}$\Msun\ $\rightarrow 10^{11.57}$\Msun). In this context it is interesting to note again that $\sim$10\% of our initial UMPEG candidates in \cite{LARGE1} were rejected when visual inspection revealed them to have two sub-\Mumpeg\ components;  these rejected candidates could represent ongoing major mergers
 that will coalesce into UMPEGs over time, consistent with the major-merger scenario of UMPEG growth. 

Finally, the inferred \zs4  mass of $M^{z=4}_{\rm UMPEG} = 10^{11.43}$\Msun\ can also be 
compared with the masses of the photometrically-selected \zs4 quiescent galaxies of \citet[][\Mstars~$=10^{11.02-11.26}$\Msun]{Kubo2018}, or with the masses of the most extreme, spectroscopically-confirmed examples (\Mstars=$10^{11.23}$\Msun, \citealt{Glazebrook2017};  \Mstars=$10^{11.49}$\Msun, \citealt{Forrest2020}).  Most of these masses
are somewhat lower than those of our projected UMPEG progenitors, but not dramatically so, and so it could be expected that 
significant numbers of
 high-mass \zs4 quiescent galaxies would be found in larger-area surveys. It seems that these observed \zs4 quiescent galaxies could potentially grow into \zs1.6 UMPEGs, and then \zs0 BCGs, through a combination of major and minor mergers. In this context we note that \cite{Kubo2018} present evidence for the growth of \zs4 massive galaxies through minor mergers, while \cite{Shi2019} report a significant number of massive (\Mstars$>10^{11}$\Msun) quiescent galaxy candidates in a \zs4 protocluster that they interpret as the central galaxies of massive (sub)haloes that one could expect to merge as the protocluster coalesces.   In further agreement with this scenario, \citet{Marsan2019} and \citet{Stockmann2020}  report that many of their ultramassive quiescent galaxies at \zs1.5--2.5 show signs of interactions and mergers.

Overall, the arguments we presented suggest a scenario in which  galaxies that follow the UMPEG evolutionary pathway may grow through major mergers at very high redshifts, $z>1.6$, before being observed as quiescent UMPEGs at \zs1.6, and then growing mildly mostly through minor mergers at lower redshifts.

\section{Conclusions}

In this paper we examined the environments of 61 ultra-massive (\Mstars $> 10^{11.5}$\Msun) galaxies that were already quiescent by \zs1.6 -- i.e., near the peak epoch of cosmic star formation.  These galaxies form the most massive part of the galaxy population at these redshifts \citep{LARGE1}, and their extremely strong clustering suggests that they are associated with dark matter halos that will grow into present-day massive galaxy clusters 
\citep{Cheema2020}.

Our analysis in this paper focused on the number of companions within 0.5~Mpc of our UMPEGs.  In their deepest part, our data allow us to detect both quiescent and star-forming companions down to \Msat$\sim$$M^*_\star$, or a mass ratio of \Msat:\Mumpeg$\sim$1:7.5; meanwhile, the wide area of the survey lets us apply a statistical fore/background correction to infer the number of physically associated 
massive
satellites.   

The main findings of our analysis are as follows: 
\begin{enumerate}

\item UMPEGs at \zs1.6 have very few companions of comparable mass:  UMPEGs have virtually no massive physical companion galaxies (mass ratios \Msat\ : \Mumpeg~$\ga$ 1:3), and on average have only one satellite per UMPEG down to mass ratio \Msat\ : \Mumpeg $\sim 1:5.5$.  

\item Given this paucity of companions, there is at present no strong evidence from galaxy counts that UMPEGs reside in overdense regions.  However, this is likely simply because our relatively shallow data do not probe sufficiently deep to find satellites in significant numbers:  in the Wide fields, where the bulk ($\sim$~90\%) of our sample resides, we are complete to \Mstars~$\approx M^*_\star \approx 10^{10.7}$\Msun\ and deeper data are needed to reach the bulk of the galaxy population at \Mstars~$\approx M^*_\star$ and below. Indeed, in the Deep fields (where we only have six UMPEGs given the small total area of these fields, but where we reach lower companion masses) such lower-mass companions are seen, with a large luminosity (or mass) gap between the UMPEG and their brightest (most massive) satellites. 

\item The presence of the luminosity (mass) gap suggests that UMPEGs reside in environments that have formed significantly before we observe them at \zs1.6.  In this scenario, dynamical friction leads to rapid merging of the most massive galaxies and leaves behind only lower-mass satellites, which 
take longer
to merge. This suggests that UMPEG progenitors may have assembled via major (mass ratio $>$ 1:4) mergers at $z>1.6$.

\item We estimate the UMPEG mass accretion rate due to mergers with satellites of mass ratio \Msat\ : \Mumpeg\ $> 1:4$ to be $\sim$13\% Gyr$^{-1}$ with a $\sim$factor-of-two systematic uncertainty.  Projected back in time, the estimated growth rate suggests that at \zs4 the typical progenitor of a \zs1.6 UMPEG had a stellar mass of $\sim10^{11.43}$\Msun\ (if that progenitor was star-forming, its \zs4 mass would have been lower).  Evolution from \zs4 to \zs1.6 could have been accomplished by major mergers, with typically a single major merger (as expected from the presence of the mass gap) with mass ratio of $\sim$2.7:1 being enough to grow the mass to the \Mumpeg $\sim10^{11.57}$\Msun\ observed at \zs1.6.  In addition to the presence of the mass gap, such early growth through major mergers is also supported by the fact that $\sim$10\% of our original UMPEG candidates are double-cored and thus possibly undergoing mergers. 

\item Were the estimated merger rate continue to \zs0, then a typical UMPEG, with \Mumpeg$\sim10^{11.6}$\Msun\ at \zs1.6 would grow to \Mstars$\sim 10^{12.0}$ through mergers with such relatively massive companions. This projected \zs0 mass is consistent with the masses of present-day massive cluster BCGs lending further support to the idea (raised by 
\citealt{Cheema2020})
that UMPEGs are the direct, already-quiescent and massive progenitors of the central galaxies of some of the present-day massive galaxy clusters such as Virgo and Coma.

\item The estimated growth rate ($\sim$13\% Gyr$^{-1}$) is relatively mild and thus is unlikely to significantly affect the shape of the steep, exponential end of the galaxy stellar mass function of passive galaxies.  Any such (minor) merger-induced change in $M^*$ will be obscured by the addition of newly-quenched massive galaxies, and the Schechter function of quiescent galaxies can be regarded as primarily a reflection of the mass-quenching process \citep{Peng2010}.

\end{enumerate}

Overall, the results presented in the present paper are consistent with the idea that \zs1.6 UMPEGs may be the direct, largely-formed progenitors of some ($\sim$1-in-8; \citealt{Cheema2020})
of the central galaxies in present-day massive clusters.  It seems plausible that galaxies that follow the UMPEG evolutionary pathway may grow through major mergers at very high redshifts, $z>1.6$ before being observed as quiescent UMPEGs at \zs1.6, and then growing mildly mostly through minor mergers at lower redshifts. 

A detailed understanding of how such progenitor galaxies already assembled and quenched by \zs1.6 remains a topic for further study, but our sample of high-$z$ UMPEGs offers opportunities for such work through follow-up spectroscopy that can constrain their metallicities and, hence assembly pathways, and through deeper imaging that can probe for evidence of merger events and search for fainter companions that could be used for halo occupation distribution (HOD) analyses.

\section*{Acknowledgments}

We thank Ivana Damjanov, Laura Parker, and Rob Thacker for their useful suggestions,  and the Natural Sciences and Engineering Research Council (NSERC) of Canada for financial support. MS is grateful to the Herzberg Astronomy and Astrophysics Research Centre for hosting his sabbatical, during which parts of this paper were written. 

This  work  is  based  on  observations  obtained  withMegaPrime/MegaCam, a joint project of CFHT and CEA/DAPNIA, at the Canada-France-Hawaii Telescope (CFHT) which is operatedby the National Research Council (NRC) of Canada, the Institut National des Science de l'Univers of the Centre National de la Recherche Scientifique (CNRS) of France, and the University of Hawaii. This work uses data products from TERAPIX and theCanadian Astronomy Data Centre. It makes use of the VIPERS-MLS database, operated at CeSAM/LAM, Marseille, France. This work is based in part on observations obtained with WIRCam, a joint project of CFHT, Taiwan, Korea, Canada and France. The research was carried out using computing resources from ACEnet and Compute Canada.


\bibliographystyle{mnras}
\bibliography{LARgE3} 










\bsp	
\label{lastpage}
\end{document}